\RequirePackage{lineno} 
\documentclass[aps,prc,superscriptaddress,showpacs,floatfix,nofootinbib,twocolumn]{revtex4}
\usepackage{graphicx} 
\usepackage{float} 
\usepackage{amssymb}
\usepackage{url}
\usepackage{harpoon}
\usepackage{amsmath}
\usepackage{multirow}
\usepackage[colorlinks,breaklinks]{hyperref}
\hypersetup{linkcolor=blue,citecolor=blue,filecolor=black,urlcolor=blue}
\newcommand{\pT} {\ensuremath{p_{\mathrm{T}}}}

\newcommand{\Dphi}{\mbox{$\Delta \phi$}}
\newcommand{\Deta}{\mbox{$\Delta \eta$}}
\newcommand{\SF}{S_{\mathrm{F}}}
\newcommand{\SB}{S_{\mathrm{B}}}
\newcommand{\ER}{\eta_{_{\mathrm{R}}}}
\newcommand{\vns}{v_{n}^{\mathrm{s}}}
\newcommand{\vnc}{v_{n}^{\mathrm{c}}}
\newcommand{\vnsr}{v_{n}^{\mathrm{s,raw}}}
\newcommand{\vncr}{v_{n}^{\mathrm{c,raw}}}
\newcommand{\vnr}{v_{n}^{\mathrm{raw}}}

\newcommand{\vtsr}{v_{2}^{\mathrm{s,raw}}}
\newcommand{\vtcr}{v_{2}^{\mathrm{c,raw}}}

\begin{document}
\title{A method for studying the rapidity fluctuation and decorrelation of harmonic flow in heavy-ion collisions}
\newcommand{\sunysb}{Department of Chemistry, Stony Brook University, Stony Brook, NY 11794, USA}
\newcommand{\bnl}{Physics Department, Brookhaven National Laboratory, Upton, NY 11796, USA}
\author{Jiangyong Jia}\email[Correspond to\ ]{jjia@bnl.gov}
\affiliation{\sunysb}\affiliation{\bnl}
\author{Peng Huo}\affiliation{\sunysb}
\begin{abstract}
An ``event-shape twist'' technique is proposed to study the longitudinal dynamics of harmonic flow, in particular the effects of rapidity fluctuation and event-plane decorrelation. This technique can distinguish between two types of rapidity decorrelation effects: a systematic rotation versus a random fluctuation of flow angles along the rapidity direction. The technique is demonstrated and the magnitude of the two decorrelation effects is predicted using the AMPT model via a single particle analysis and two-particle correlation analysis. An observed decorrelation can be attributed to a systematic rotation of event-plane angle along the pseudorapidity, consistent with a collective response to an initial state twist of the fireball proposed by Bozek {\it et.al.}. This rotation is also observed for several higher-order harmonics with the same sign and similar magnitudes. 
\end{abstract}
 \maketitle

\section{Introduction} 
High energy heavy ion collisions at the RHIC and the LHC have created a new form of nuclear matter composed of deconfined and yet strongly interacting quarks and gluons. This matter exhibits significant azimuthal anisotropy in its particle production in the transverse plane~\cite{Adams:2004bi,Adare:2010ux,ALICE:2011ab,CMS:2012wg,Aad:2012bu}. Such anisotropy is the result of a collective response of the system to the asymmetric collision geometry in the initial state, and is well described by relativistic viscous hydrodynamic models~\cite{Teaney:2009qa}. The particle distribution in azimuthal angle $\phi$ is usually expressed in terms of a Fourier series:
\begin{equation}
\label{eq:flow0}
\frac{dN}{d\phi}\propto1+2\sum_{n=1}^{\infty}v_{n}\cos n(\phi-\Phi_{n})\;,
\end{equation}
where $v_n$ and $\Phi_n$ (event-plane angle, EP or flow angle) represent the magnitude and phase of the $n^{\mathrm{th}}$-order flow harmonic. Initially, these flow harmonics were attributed to various shape components of the initial geometry, whose magnitudes and directions can fluctuate strongly event-by-event (EbyE), leading to large EbyE fluctuation of $v_n$ and $\Phi_{n}$~\cite{Alver:2010gr,Bhalerao:2011yg}. However, recent measurement of event-plane correlation~\cite{Jia:2012sa,Aad:2014fla} and theoretical calculations~\cite{Teaney:2010vd,Qiu:2011iv,Gardim:2011xv,Teaney:2012ke,Qiu:2012uy} show that the flow harmonics are also strongly modified by non-linear mode-mixing effects in the final state. A central focus of current research is to understand various types of fluctuations in the initial state and later time, and how these fluctuations influence the hydrodynamic evolution of the matter in the final state. 

Experimentally, the flow coefficients $v_n$ are also obtained by assuming a factorization of the Fourier coefficients of two-particle angular correlation (2PC) into the product of single-particle flow coefficients $v_n$~\cite{Adcox:2002ms,Aamodt:2011by,CMS:2012wg,Aad:2012bu}:
\begin{equation}
\label{eq:flow2}
v_{n,n}(\pT^{\rm a},\eta^{\rm a},\pT^{\rm b},\eta^{\rm b})= v_{n}(\pT^{\rm a},\eta^{\rm a})v_n(\pT^{\rm b},\eta^{\rm b})\;,
\end{equation}
where coefficients $v_{n,n}$ are obtained by 2PC analysis in relative azimuthal angle $\Dphi=\phi^{\rm a}-\phi^{\rm b}$ for particle ``a'' at $\pT^{\rm a}$ and $\eta^{\rm a}$ and particle ``b'' at $\pT^{\rm b}$ and $\eta^{\rm b}$:
\begin{equation}
\label{eq:flow3}
\frac{dN_{\mathrm{pairs}}}{d\Dphi}\propto1+2\sum_{n=1}^{\infty}v_{n,n}(\pT^{\rm a},\eta^{\rm a},\pT^{\rm b},\eta^{\rm b})\cos n(\Dphi)\;.
\end{equation}
The factorization relation works as long as flow angles $\Phi_{n}$ are independent of $\pT$ and $\eta$.

The information of the EbyE fluctuations is generally described by the full probability density distribution in terms of $v_n$ and $\Phi_n$~\cite{Huo:2013qma}:
\begin{equation}
\label{eq:flow1}
p(v_n,v_m,...., \Phi_n, \Phi_m, ....)=\frac{1}{N_{\mathrm{evts}}}\frac{dN_{\mathrm{evts}}}{dv_ndv_m...d\Phi_{n}d\Phi_{m}...}\;.
\end{equation}
Initial measurements of a subset of these flow observables, namely $p(v_n)$~\cite{Aad:2013xma} and event-plane correlations $p(\Phi_n,\Phi_m....)$~\cite{Jia:2012sa,Aad:2014fla}, have been performed by the LHC experiments. The measured event-plane correlations are reproduced by EbyE hydrodynamics~\cite{Qiu:2012uy,Teaney:2012gu} and AMPT transport model~\cite{Bhalerao:2013ina} calculations, providing valuable insights on the linear and non-linear effects in the collective evolution. Additional observables, such as correlation between $v_n$ and $v_m$ ($p(v_n,v_m)$), can be further explored~\cite{Huo:2013qma} using the  event-shape selection technique~\cite{Schukraft:2012ah,Huo:2013qma} or cumulant method~\cite{Bilandzic:2013kga}. 

Flow fluctuations not only occur across different events, but also occur within the same event~\cite{Gardim:2012im,Heinz:2013bua}. Due to the presence of quantum fluctuations, non-linear effects and initial flow, flow angles $\Phi_n$ can fluctuate as a function of transverse momentum $\pT$ or pseudorapidity $\eta$. These intra-event fluctuations can break the factorization relation Eq.~\ref{eq:flow2}. The flow angle fluctuation and the breaking of the factorization in $\pT$ space have been studied extensively. However, experimental study of flow fluctuations in the longitudinal direction is limited, since most experimental methods assume flow angle to be independent of $\eta$. Nevertheless, theoretical studies based on either an EbyE hydrodynamic model~\cite{Bozek:2010vz,Petersen:2011fp} or a transport model~\cite{Xiao:2012uw} have shown that the correlation between flow angles in two $\eta$ regions decreases with their $\eta$ separation. This decorrelation effect was also explored by the event-shape selection technique in Ref.~\cite{Huo:2013qma}: events selected with smaller or larger $v_m$ in very forward $\eta$ exhibit a strong forward/backward (FB) asymmetry of $v_m$ near mid-rapidity, and this asymmetry also feeds to other flow harmonics $v_n$ ($n\neq m$) via non-linear effects~\cite{Huo:2013qma}. 

\begin{figure}[h!]
\centering
\includegraphics[width=1\linewidth]{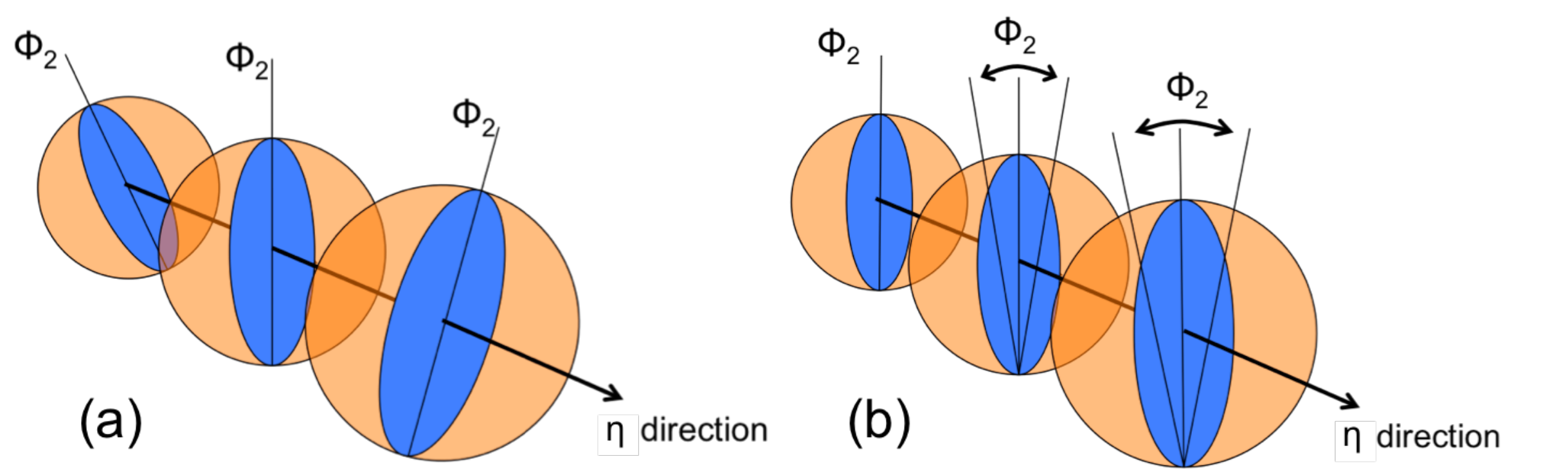}
\caption{\label{fig:1} (Color online) The two scenarios for the rapidity fluctuation of $v_2$: (a) the fluctuation arises from a systematic rotation as a function of $\eta$~\cite{hanna}, (b) the fluctuation is random between different rapidity ranges.}
\end{figure}

One possible explanation for the event-plane decorrelation is based on the ``torqued fireball'' idea, proposed by Bozek {\it et.al.}~\cite{Bozek:2010vz} (see Fig.~\ref{fig:1}(a)). The idea can be explained briefly below (with some generalization): Particles in the forward (backward) rapidity is preferably produced by the participants in the forward-going (backward-going) nucleus (responsible also for the FB-asymmetry of the multiplicity distribution in p+A collisions). Since the shape and the orientation of the participating-part of the two colliding nuclei fluctuate semi-independently, the shape of the fireball in forward $\eta$ should be more similar to that of the participants in the forward-going nucleus and vice versa. In other words, if one calculate the eccentricity $\epsilon_m$ and participant-plane angle $\Phi_m^{*}$ separately for the two nuclei (labeled by the subscript F and B), then we expect the orientation of the initial fireball along $\eta$ to interpolate between $(\Phi_m^{*})_{\rm F}$ and $(\Phi_m^{*})_{\rm B}$. The hydrodynamic expansion of this torqued fireball leads to a torqued collective flow, resulting in the systematic rotation of the flow angle. This is a generic initial state long-range effect, which is naturally included in the AMPT transport model~\cite{Lin:2004en}. The authors also proposed a cumulant method to measure this rotation, but the expected signal is rather small once averaged over many events. 

In this paper, we propose an experimental method with increased sensitivity to the longitudinal flow fluctuation and decorrelation effects. This method can distinguish between a continuous rotation of the flow angle with $\eta$ from a random fluctuation from one $\eta$ region to the next region (see Fig.~\ref{fig:1}(b)). Our method is based on a simple procedure called ``event-shape twist''. A twist of the $m^{\mathrm{th}}$-order EP angle between the forward and backward reference pseudorapidity regions, $\Delta\Phi_m$, is calculated event-by-event. Events are then divided in ranges of $\Delta\Phi_m$, and within each class, the $n^{\mathrm{th}}$-order flow angle $\Phi_n$ is then calculated as a function of $\eta$. If the process contributing to Fig.~\ref{fig:1}(a) is significant, one expects to observe a gradual rotation of $\Phi_n$ with $\eta$ in the same direction as $\Delta\Phi_m$. This procedure preferably selects events with large twist angle in a particular direction, so the resulting signal is easier to measure. We show two implementations of the method, based on either the single particle distribution or two-particle correlations. The AMPT model~\cite{Lin:2004en} is used to validate these implementations, as well as to provide predictions that can be compared to experimental data.

\section{The single particle method}
\label{sec:1}

The AMPT (``A Muti-Phase Transport'') model~\cite{Lin:2004en} has been used to study the harmonic flow~\cite{Xu:2011jm,Xu:2011fe,Ma:2010dv}. It combines the initial fluctuating geometry based on Glauber model from HIJING and final state interaction via a parton and hadron transport model, with the collective flow generated mainly by the partonic transport. The initial condition of the AMPT model contains significant longitudinal fluctuations that can influence the collective dynamics~\cite{Pang:2012he,Pang:2012uw,Huo:2013qma,Pang:2013pma}. The model simulation is performed with string-melting mode with a total partonic cross-section of 1.5 mb and strong coupling constant of $\alpha_{\rm s}=0.33$~\cite{Xu:2011fe}. This setup has been shown to reproduce the experimental $\pT$ spectra and $v_n$ data at RHIC and the LHC~\cite{Xu:2011fe,Xu:2011fi}. 

\begin{figure}[h!]
\centering
\includegraphics[width=1\linewidth]{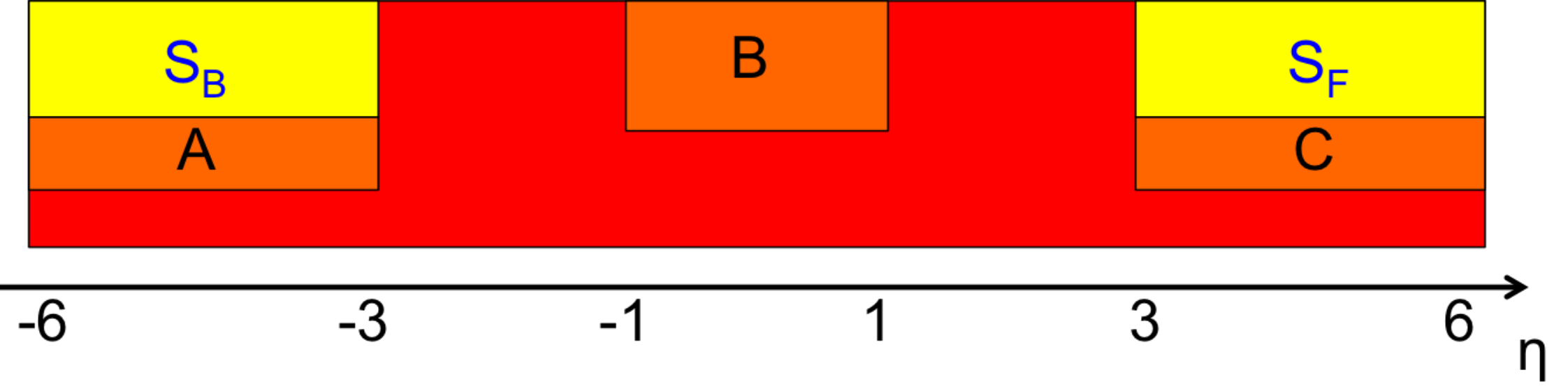}
\caption{\label{fig:2} (Color online) The $\eta$-ranges of the subevents for the event-shape twist ($\SB$ and $\SF$) and for calculating the reference event-plane angles via Eqs.~\ref{eq:rot1}-\ref{eq:rot3} (A,B and C). Note that subevent $\SB$ (A) or $\SF$ (C) contains half (a quarter) of the particles randomly selected from $-6<\eta<-3$ or $3<\eta<6$, and subevent B contains half of the particles randomly selected from $-1<\eta<1$. The subevents $\SB$ and $\SF$ together is also denoted as subevent S.}
\end{figure}

The AMPT data used in this study is generated for $b=8$~fm Pb+Pb collisions at LHC energy of $\sqrt{s_{\mathrm{NN}}}=2.76$ TeV, corresponding to $\sim 30\%$ centrality. The particles in each event are divided into subevents along $\eta$ as shown in Fig.~\ref{fig:2}. There are five independent subevents labelled as $\SB$, $\SF$, A, B and C, together with subevent S obtained by combining $\SB$ and $\SF$,  are used in the analysis. Note that one half or one quarter of the particles in $-6<\eta<-3$ ($3<\eta<6$) are randomly selected for subevents $\SB$ ($\SF$) or A (C), respectively. Furthermore, the particles in subevents $\SB$ and $\SF$ are used only for event-shape selection, and are excluded for the $v_n$ calculation. This choice of subevents and analysis scheme ensures that the event-shape selection does not introduce non-physical biases to the $v_n$ measurements.

The flow vector in each subevent is calculated as:
\begin{eqnarray}
\nonumber
&&\overrightharp{q}_n =(q_{\rm x,n},q_{\rm y,n}) = \frac{1}{\Sigma w}\left(\textstyle\Sigma (w\cos n\phi), \Sigma (w\sin n\phi)\right)\;, \\\label{eq:me1}
&&\tan n\Psi_n = \frac{q_{\rm y,n}}{q_{\rm x,n}}\;,
\end{eqnarray}
where the weight $w$ is chosen as the $\pT$ of each particle and $\Psi_n$ is the measured event-plane angle. Due to finite number effects, $\Psi_n$  smears around the true event-plane angle $\Phi_n$. In the limit of infinite particle multiplicity, the magnitude of the flow vector defined this way is equal to the weighted average of $v_n$: $(v_n)_w=\textstyle\Sigma w v_n/ \textstyle\Sigma w$. In this study, each subevent in Fig.~\ref{fig:2} has 1000-2000 particles, so $q_n$ is expected to follow closely the $(v_n)_w$.

For each generated event, the $q_n$ and $\Psi_n$ with $n=2$--5 are calculated for the five independent subevents and subevent S, a total of 48 quantities. For clarity, we shall use subscript ``$m$'' to denote the $m^{\mathrm{th}}$-order flow vectors from subevent S,  and use the subscript ``$n$'' to denote the $n^{\mathrm{th}}$-order harmonic flow calculated in the rest of the event. The event-shape selections are performed for $m=2$ and 3 by dividing the generated events into 10 bins in $q_m^{\mathrm{S}}$ with equal statistics. Events in each $q_m^{\mathrm{S}}$ bin is further divided into 10 bins with equal statistics based on the relative EP angles between $\SB$ and $\SF$ (hence the word ``twist''):
\begin{eqnarray}
\Psi_m^{\mathrm {cut}} = m\left(\Psi_m(\SF)-\Psi_m(\SB)\right)
\end{eqnarray}
A significant part of the relative spread between $\Psi_m(\SB)$ and $\Psi_m(\SF)$ reflects the random smearing due to finite multiplicity in $\SB$ and $\SF$. This random smearing effect is expected to be particularly larger for small $q_m^{\mathrm{S}}$ bin, where the corresponding $v_m$ signal is small. Hence it is necessary to select both $q_m^{\mathrm{S}}$ and $\Psi_m^{\mathrm {cut}}$.

\begin{figure}[h!]
\centering
\includegraphics[width=1\linewidth]{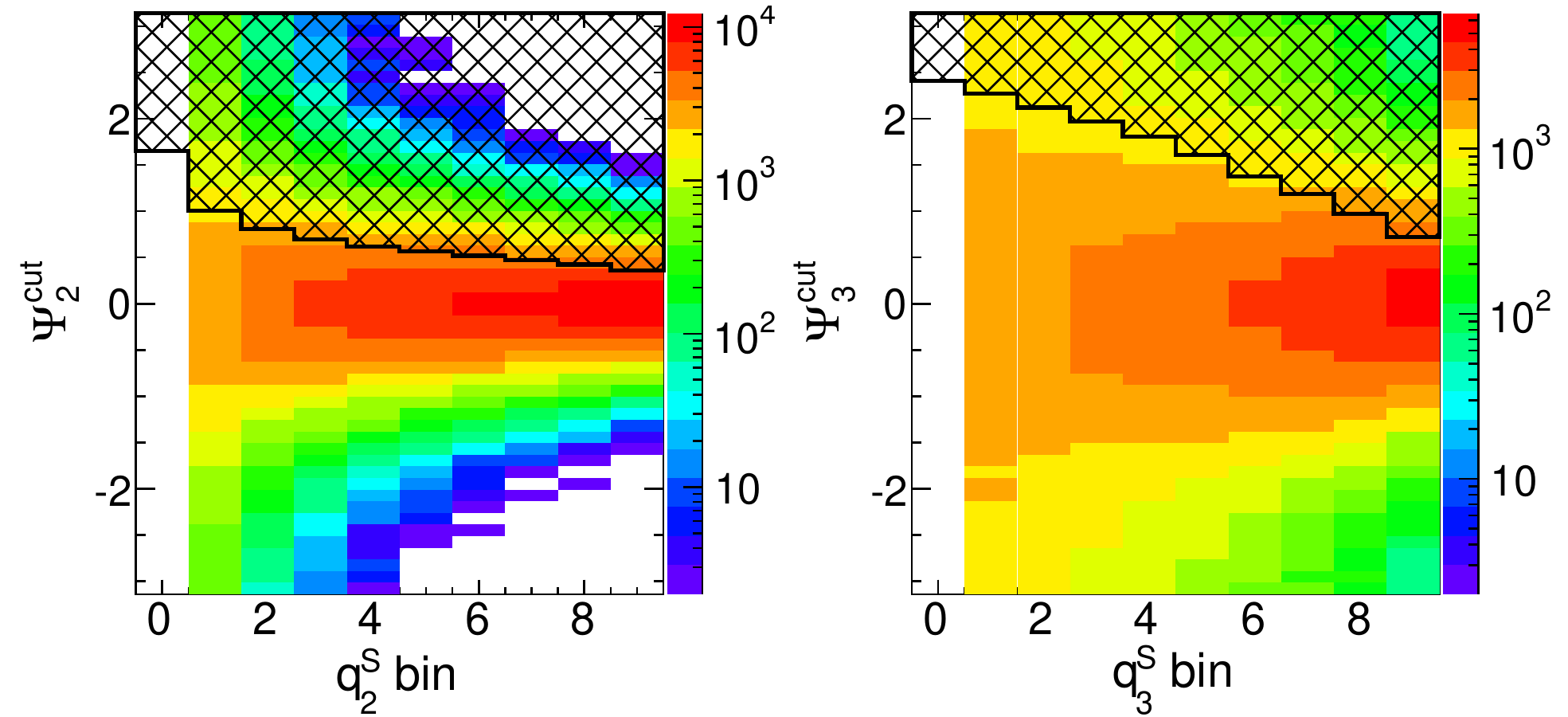}
\caption{\label{fig:3} (Color online) The $\Psi_m^{\mathrm {cut}}$ distribution in ten $q_m^{\mathrm{S}}$ bins with equal statistics for $m=2$ (left panel) and $m=3$ (right panel). The hashed area represents the 10\% of the events with largest $\Psi_m^{\mathrm {cut}}$ in each $q_m^{\mathrm{S}}$ bin, and only these events are used in this paper.}
\end{figure}

Figure~\ref{fig:3} shows the performance of the event-shape selection on $q_m^{\mathrm{S}}$ and $\Psi_m^{\mathrm {cut}}$ for $m=2$ (left) and 3 (right). In general the $\Psi_m^{\mathrm {cut}}$ distribution in small $q_m^{\mathrm{S}}$ bin is broader as expected. The $\Psi_m^{\mathrm {cut}}$ distribution is also tighter for $m=2$ than that for $m=3$, reflecting the fact that $v_2>v_3$ and hence a smaller random smearing effect for $m=2$. The hashed area selects, in each $q_m^{\mathrm{S}}$ bin, the 10\% of the events with largest $\Psi_m^{\mathrm {cut}}$ values. The results in this paper are obtained entirely from these ten event classes. Hence they are conveniently referred to by their bin number in $q_m^{\mathrm{S}}$, with the understanding that each refers to events with top 10\% of $\Psi_m^{\mathrm {cut}}$ values in each $q_m^{\mathrm{S}}$ bin. 

To investigate the rapidity fluctuation, Fourier coefficients for particles at $\eta$ relative to the EP angle in a reference subevent R$\in$\{A, B,C\} can be expressed as:
\begin{eqnarray}
\nonumber
\vnc (\eta) &=& \langle\cos n\left(\phi(\eta)-\Phi_n(\ER)\right)\rangle\\\label{eq:rot1}
\vns (\eta) &=& \langle\sin n\left(\phi(\eta)-\Phi_n(\ER)\right)\rangle
\end{eqnarray}
where the average is over all particles at $\eta$, then over the events. If the true event-plane angle $\Phi_n$ is independent of $\eta$, then $\vns=0$ and $\vnc=v_n$. However, if the $\Phi_n$ angle rotates along $\eta$, then $\vns$ may not be zero. The change of the EP angle at $\eta$ relative to that in R, $\Delta\Phi_n^{\mathrm{rot}}$, can be expressed as:
\begin{eqnarray}
\label{eq:rot2}\tan (n\Delta\Phi_n^{\mathrm{rot}}) =\frac{\langle\sin n\left(\Phi_n(\eta)-\Phi_n(\ER)\right)\rangle}{\langle\cos n\left(\Phi_n(\eta)-\Phi_n(\ER)\right)\rangle}= \frac{\vns}{\vnc}
\end{eqnarray}
where the average is performed over the events. This relation can also be obtained from the raw EP angle and raw Fourier coefficients:
\begin{eqnarray}
\nonumber
\hspace*{-0.5cm}\tan (n\Delta\Phi_n^{\mathrm{rot}}) &=&\frac{\langle\sin n\left(\phi(\eta)-\Phi_n(\ER)\right)\rangle}{\langle\cos n\left(\phi(\eta)-\Phi_n(\ER)\right)\rangle} \\\label{eq:rot3}
\hspace*{-0.5cm}&=& \frac{\langle\sin n\left(\phi(\eta)-\Psi_n(\ER)\right)\rangle}{\langle\cos n\left(\phi(\eta)-\Psi_n(\ER)\right)\rangle} = \frac{v_n^{\mathrm{s,raw}}}{v_n^{\mathrm{c,raw}}}
\end{eqnarray}
where we have used the fact that the smearing of $\Psi_n$ around $\Phi_n$ cancels out in the ratio, and hence no EP resolution correction is needed in calculating the rotation angle $n\Delta\Phi_n^{\mathrm{rot}}$. The next section presents result of $n\Delta\Phi_n^{\mathrm{rot}}$ ($n=2$--5) for each event class selected by the $q_m^{\mathrm{S}}$ and $\Psi_m^{\mathrm {cut}}$ ($m=2$ and 3) in the hashed regions of Fig.~\ref{fig:3}.

\section{Results from the single particle method}
\label{sec:3}
\begin{figure}[h!]
\centering
\includegraphics[width=1\linewidth]{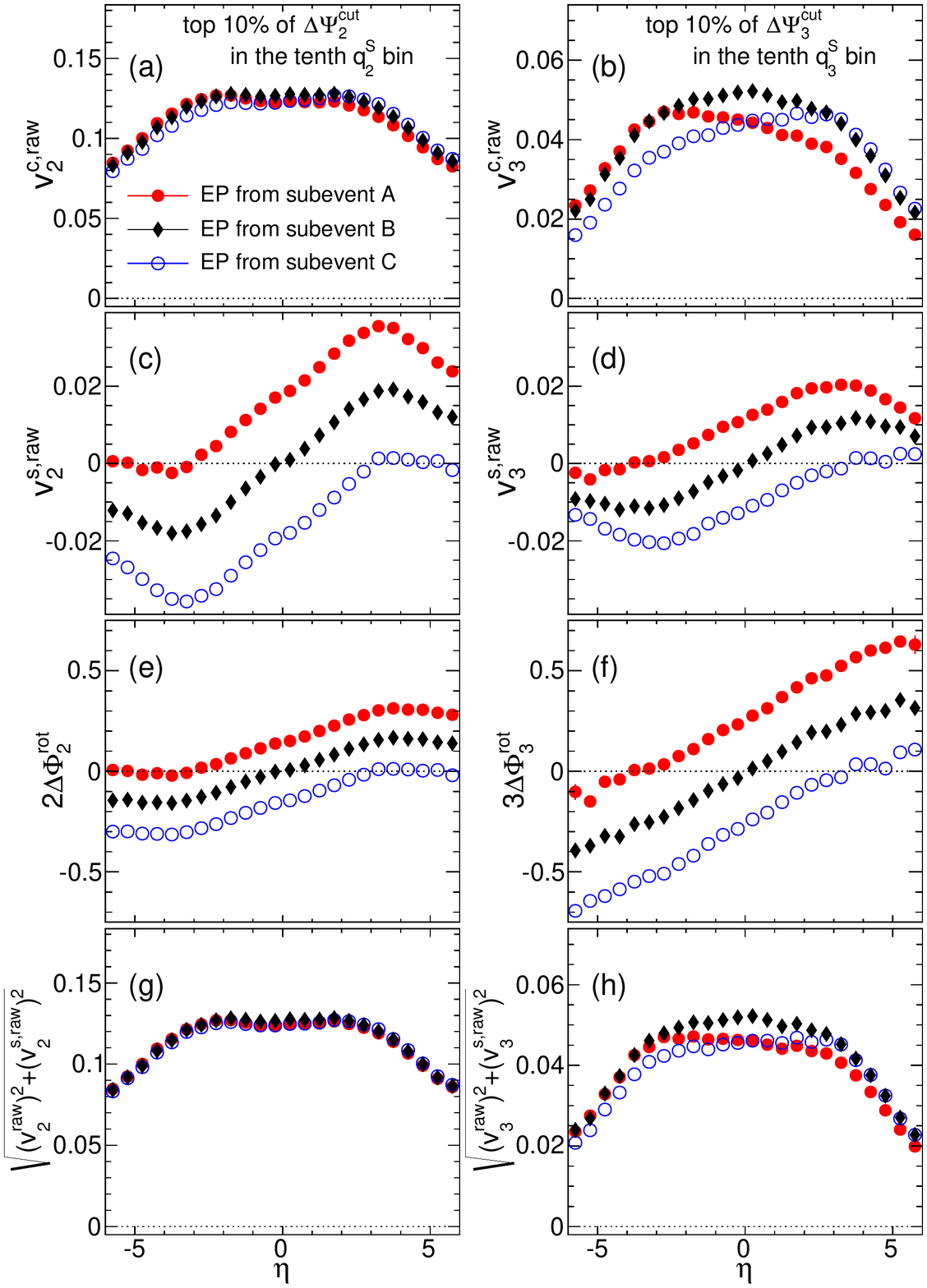}
\caption{\label{fig:4} The $\vncr(\eta)$ (top row), $\vnsr(\eta)$ (second row), rotation angle $n\Delta\Phi_n^{\mathrm{rot}}$ (third row) and $\sqrt{(\vncr)^2+(\vnsr)^2}$ (bottom row) relative to reference EP angle calculated in subevents A, B and C in Eqs.~\ref{eq:rot1}-\ref{eq:rot3}. They are obtained via Eq.~\ref{eq:rot3} for events selected with largest $\Psi_m^{\mathrm {cut}}$ in the tenth $q_m^{\rm S}$ bin for $m=2$ (left column) and $m=3$ (right column).}
\end{figure}

The left column of Fig.~\ref{fig:4} shows the $\vtcr(\eta)$ and $\vtsr(\eta)$ values as well as the rotation angle $2\Delta\Phi_2^{\mathrm{rot}}$ relative to the EP calculated in subevent A, B or C. They are obtained via Eq.~\ref{eq:rot3} for events in the largest $q_2^{\rm s}$ bin and with largest $\Psi_2^{\mathrm {cut}}$ values. A significant nonzero $\vtsr(\eta)$ is observed, which varies linearly with $\eta$; this suggests a systematic rotation of the EP angle as a function of $\eta$ for the selected events. The general trends of the $\Delta\Phi_2^{\mathrm{rot}}(\eta)$ are nearly identical for the three reference subevents. Similar behavior is also observed for $v_{3}^{\mathrm{s,raw}}(\eta)$ for events selected with largest $q_3^{\rm s}$ and $\Psi_3^{\mathrm {cut}}$, as shown in the right column of Fig.~\ref{fig:4}.

 The behavior of $\vnsr$ shown in Fig.~\ref{fig:4} suggests that the previously observed rapidity decorrelation of the $v_2$ or $v_3$ can be attributed, at least partially, to a systematic rotation of their EP angles in $\eta$. Clearly, this rotation would also break the factorization of the 2PC $v_{n,n}$ to the single particle $v_n$ in different rapidity, i.e. $v_{n,n}(\eta^{\rm a},\eta^{\rm b})\neq v_n(\eta^{\rm a}) v_n(\eta^{\rm b})$. In this case the $\vnc$ calculated from the EP method no longer correctly represents the true flow coefficient. Instead, the raw $v_n$ signal can be estimated by $\vnr= \sqrt{(\vncr)^2+(\vnsr)^2}$. The results of this estimation are plotted in the bottom row of Fig.~\ref{fig:4}, which shows a much smaller forward/backward asymmetry than the $\vncr(\eta)$. The small residual asymmetry, more obvious for $n=3$, may reflect the contribution of the random component of the rapidity fluctuation (see Fig.~\ref{fig:1}(b)). 

\begin{figure}[h!]
\centering
\includegraphics[width=1\linewidth]{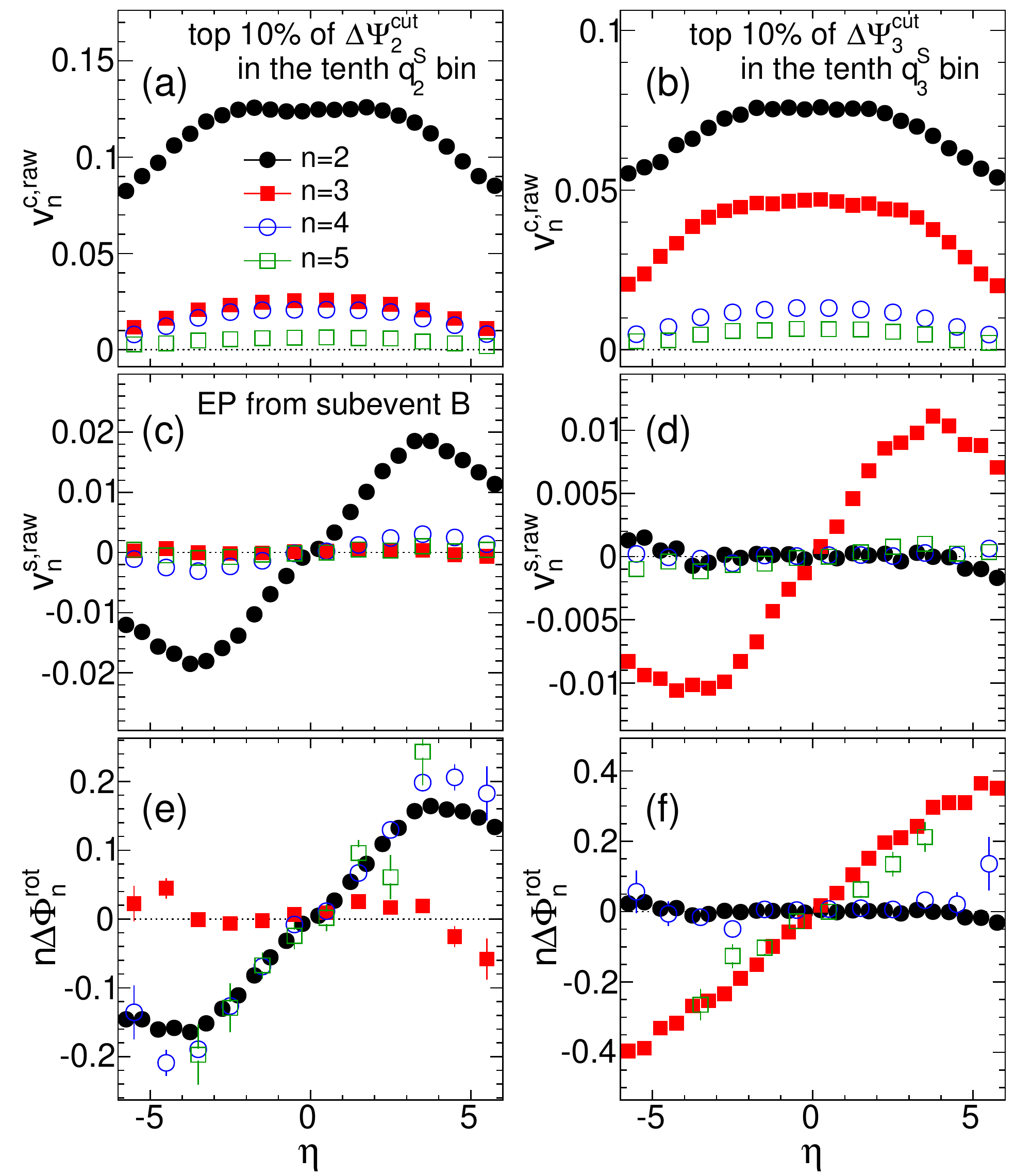}
\caption{\label{fig:5} The $\vncr$ (top row), $\vnsr$ (middle row) and the rotation angle $n\Delta\Phi_n^{\mathrm{rot}}$ (bottom row) for events selected with largest $\Psi_m^{\mathrm {cut}}$ in the tenth $q_m^{\rm S}$ bin, where $m=2$ (left column) and $m=3$ (right column) and $n=2$--5. The reference event-plane angle used in Eqs.~\ref{eq:rot1}-\ref{eq:rot3} is calculated in subevent B.}
\end{figure}

One important issue in the study of flow is the final state non-linear effects, which mixes between harmonics of different order. Previous event-plane correlation~\cite{Bhalerao:2013ina} and event-shape selection studies~\cite{Huo:2013qma} revealed a strong non-linear coupling between harmonics of different order in the AMPT model. It is natural to ask whether the event-plane rotation effects shown in Fig.~\ref{fig:4} also feed into the higher-order harmonics via these non-linear coupling effects. In order to check this, the $\vncr(\eta)$, $\vnsr(\eta)$ and $n\Delta\Phi_n^{\mathrm{rot}}$ for higher-order harmonics are calculated for events used in Fig.~\ref{fig:4}. The results obtained with reference subevent B via Eq.~\ref{eq:rot3} (results from subevents A and C are similar) are shown in Fig.~\ref{fig:5} for $m=2$ (left column) and $m=3$ (right column). Clear nonzero, rapidity-odd $\vnsr(\eta)$ distributions are observed for several higher-order harmonics with $n>m$. The magnitudes of $\vncr(\eta)$ and $\vnsr(\eta)$ both drop rapidly with increasing $n$, resulting in $n\Delta\Phi_n^{\mathrm{rot}}$ values (from Eq.~\ref{eq:rot2}) that change more slowly with $n$. Systematic rotations are also observed for several higher-order harmonics: For the event-shape twist based on elliptic flow ($m=2$), significant rotations are observed for $n=4$ and 5 with a magnitude similar to that for $n=2$, reflecting a non-linear coupling of higher-order harmonics to the lower-order harmonics. Similarly for event-shape twist based on triangular flow ($m=3$), the rotation for $n=5$ is coupled to $n=3$.

\begin{figure*}[t!]
\centering
\includegraphics[width=1\linewidth]{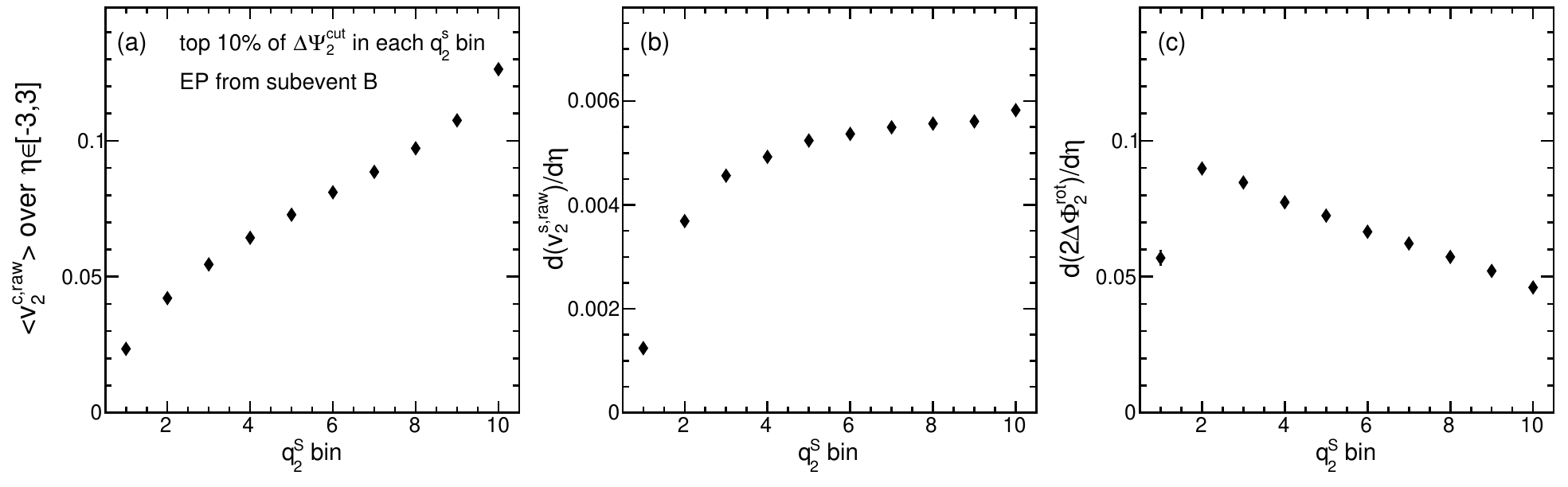}
\caption{\label{fig:6} The $\left\langle\vtcr\right\rangle$ (left), $d(\vtsr)/d\eta$ (middle) and $d(2\Delta\Phi_2^{\mathrm{rot}})/d\eta$ (right) as a function of $q_2^{\rm S}$ bin defined in the hashed region in the left panel of Fig.~\ref{fig:3}. The reference event-plane angle used in Eqs.~\ref{eq:rot1}-\ref{eq:rot3} is calculated in subevent B.}
\end{figure*}

The same procedure is repeated for other nine $q_{m}^{\rm S}$ bins marked in the hashed region of Fig.~\ref{fig:3}. The $\vnsr$ and $n\Delta\Phi_n^{\mathrm{rot}}$ are found to vary linearly with $\eta$ and they all cross zero at $\eta=0$ similar to what is shown in the middle and third rows of Fig.~\ref{fig:4}. The rate of this rotation thus can be quantified by a linear fit to $\vnsr$ and $n\Delta\Phi_n^{\mathrm{rot}}$ over the region $-3<\eta<3$ for each $q_{m}^{\rm S}$ bin. Note that since the rotation angle is small at mid-rapidity and $\vncr$ changes slowly with $\eta$ (see the top row of Fig.~\ref{fig:4}), we obtain the following relations for events selected on $q_{m}^{\rm S}$ and $\Psi_m^{\mathrm {cut}}$:
\begin{eqnarray}
\nonumber
n\Delta\Phi_n^{\mathrm{rot}} \approx \tan (n\Delta\Phi_n^{\mathrm{rot}})&=&\frac{\vnsr(\eta)}{\vncr(\eta)}\approx\frac{\vnsr(\eta)}{\left\langle \vncr\right\rangle},\\\label{eq:rot4}
\kappa_{m,n} = \frac{d(n\Delta\Phi_n^{\mathrm{rot}})}{d\eta} &\approx& \frac{d(\vnsr)}{d\eta}\frac{1}{\left\langle \vncr\right\rangle}
\end{eqnarray}
where $\left\langle \vncr\right\rangle$ is the $\vncr$ value averaged over \mbox{$-3<\eta<3$}.

Figure~\ref{fig:6} shows the values of $\left\langle \vtcr\right\rangle$, $d(\vnsr)/d\eta$ and $d(n\Delta\Phi_n^{\mathrm{rot}})/d\eta$ as a function of $q_2^{\rm S}$ bin. The value of $\left\langle \vtcr\right\rangle$ increases continuously for larger $q_2^{\rm S}$ bin which has a large $v_2$. The slope of $\vtsr$ quickly saturates past the fourth $q_2^{\rm S}$ bin, suggesting that the component responsible for rapidity decorrelation is nearly independent of the values of $q_2^{\rm S}$ (or $v_2$) of the events. Naturally, these behaviors lead to a decrease of rate of rotation $d(2\Delta\Phi_2^{\mathrm{rot}})/d\eta$ for larger $q_2^{\rm S}$ bin as shown in the Fig.~\ref{fig:6}(c). 

\section{Two-particle correlation method and results}
\label{sec:5}
\begin{figure}[h]
\centering
\includegraphics[width=0.8\linewidth]{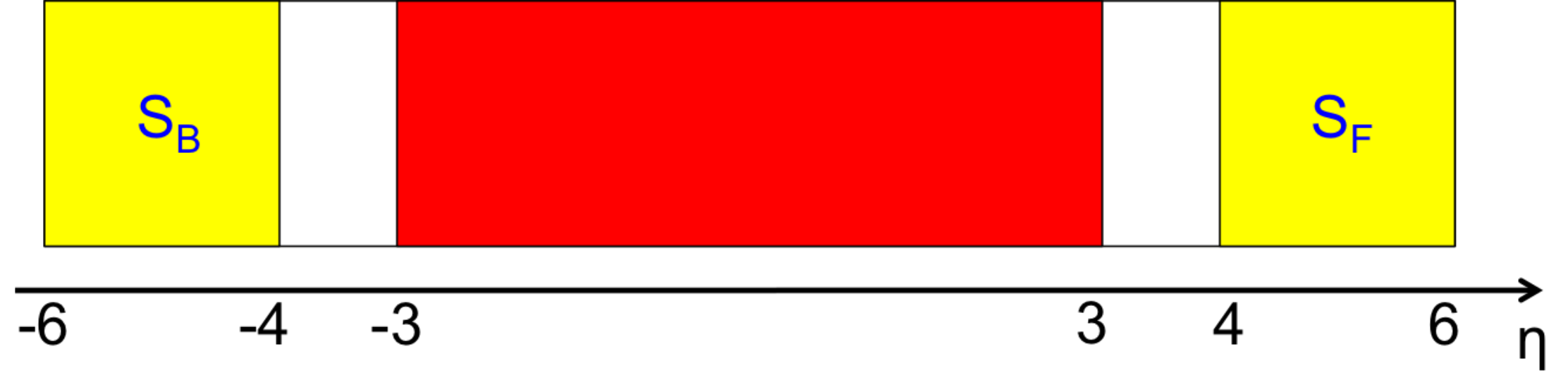}
\caption{\label{fig:7}  (Color online) The $\eta$-range of the subevents for the event-shape twist ($\SB$ and $\SF$ over $-6<\eta<-4$ or $4<\eta<6$ ) and the particles used for two-particle correlation analysis ($-3<\eta<3$). The subevents $\SB$ and $\SF$ together is also denoted as subevent S.}
\end{figure}

The fact that the amount of rotation is a linear function of $\eta$ suggests that this rotation can also be extracted easily with the two-particle correlation method. In this method, the definition of the subevents can be simplified as illustrated by Fig.~\ref{fig:7}. The partition of $q_{m}^{\rm S}$ bins is identical to those shown in Fig.~\ref{fig:2}, except that the definition of subevent S is different. For each event class, the correlation function is obtained as the ratio of the same-event distribution (S) to the mixed-event distribution (B)~\cite{Adare:2008ae}:
\begin{eqnarray}
\label{eq:2pc1}
C(\Dphi,\Deta) = \frac{S(\Dphi,\Deta)}{B(\Dphi,\Deta)}
\end{eqnarray}
where same-event distribution is obtained from pairs of particles in the same events, and mixed-event distribution is obtained by pairing particles from different events to account for triangle structure in $\Deta$ due to $|\eta^{\rm a},\eta^{\rm b}|<3$. 

In previous studies, the event-plane angle $\Phi_n$ is assumed to be independent of the $\eta$, and the Fourier coefficients of the correlation function can be factorized into $v_n$ of the two particles: $C(\Dphi,\Deta)\propto 1+\Sigma v_n^{\rm a} v_n^{\rm b}\cos n\Dphi$. However, if the event-plane angle rotates linearly in $\eta$, then the formula need to be modified:
\begin{eqnarray}
\nonumber
C(\Dphi,\Deta) &\propto& 1+2\sum v_n^{\rm a} v_n^{\rm b}\cos \left(n\Dphi-n\Delta\Phi_n^{\mathrm{rot}}\right)\\\nonumber
&\approx& 1+2\sum v_n^{\rm a} v_n^{\rm b}\cos \left(n\Dphi-\kappa_{m,n}\Deta\right)\\\label{eq:2pc2}
\end{eqnarray}
where $n\Delta\Phi_n^{\mathrm{rot}} \approx \kappa_{m,n}\Deta$ accounts for the rotation of event-plane angle for pairs separated by $\Deta$ in a given $q_{m}^{\rm S}$ and $\Psi_m^{\mathrm {cut}}$ event class. A nonzero value of $\kappa_{m,n}$ leads to a phase shift that increase with $|\Deta|$, and the correlation function is no longer an even function in $\Dphi$. 
\begin{figure*}[t!]
\centering
\includegraphics[width=1\linewidth]{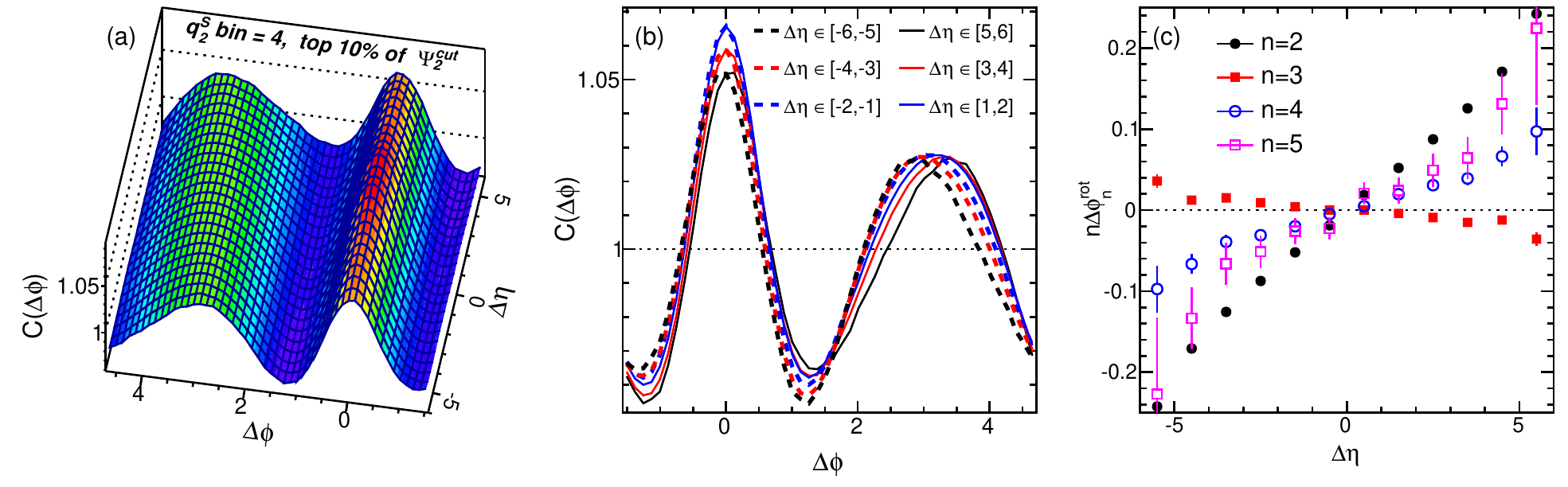}
\caption{\label{fig:8} The 2D correlation function (left), 1D correlation function in different $\Deta$ slices (middle panel) and the extracted $n\Phi_n^{\mathrm{rot}}$ for $n=2$--5 (right), for events selected in the fourth $q_2^{\rm S}$ bin with largest $\Psi_2^{\mathrm {cut}}$ values.}
\end{figure*}
\begin{figure*}[t!]
\centering
\includegraphics[width=1\linewidth]{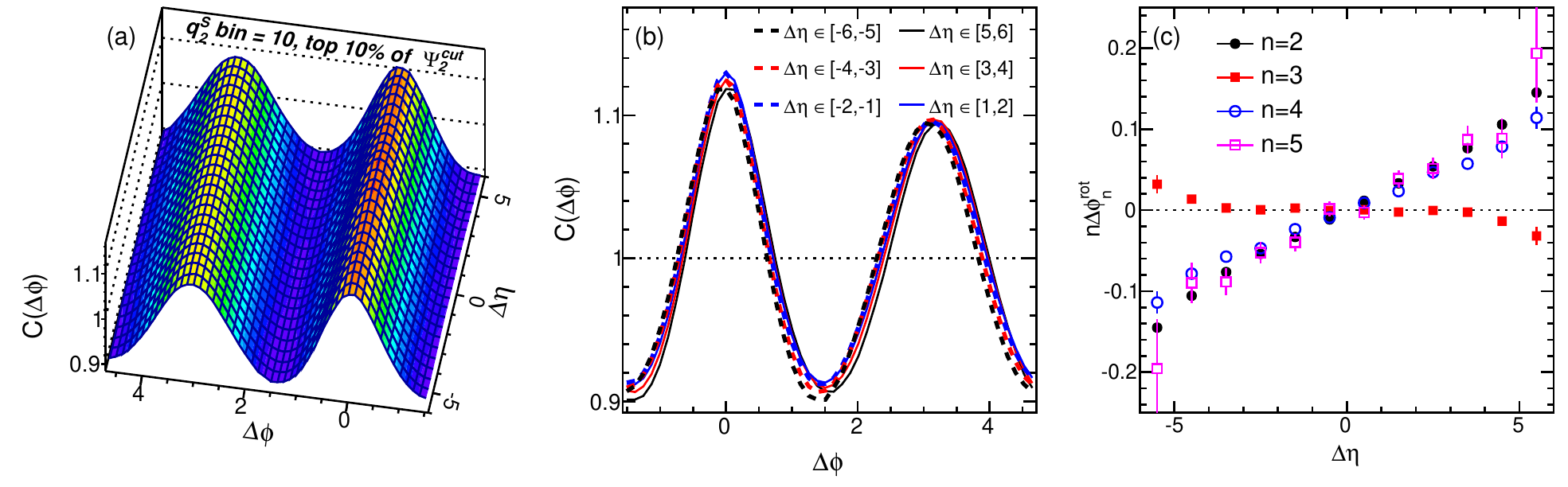}
\caption{\label{fig:9} The 2D correlation function (left), 1D correlation function in different $\Deta$ slices (middle panel) and the extracted $n\Phi_n^{\mathrm{rot}}$ for $n=2$--5 (right), for events selected in the tenth $q_2^{\rm S}$ bin with largest $\Psi_2^{\mathrm {cut}}$ values.}
\end{figure*}

We illustrate the results obtained for event-shape selection based on $m=2$. Figures~\ref{fig:8} and \ref{fig:9} summarize the results of 2PC obtained for the fourth $q_2^{\rm S}$ bin and the tenth $q_2^{\rm S}$ bin, respectively. A clear phase shift is seen in both the 2D correlation function and 1D correlation function projected in various $\Deta$ slice. The rotation (Fig.~\ref{fig:8}(c) and Fig.~\ref{fig:9}(c)) is indeed a linear function of $\Deta$ for all $n=2-5$, and rotation for $n=4$ and 5 is correlated with that for $n=2$. On the other hand, the rotation for $n=3$ is small and slightly anti-correlated with $n=2$. These observations are qualitatively consistent with the results shown in Fig.~\ref{fig:5}.

\section{Discussion and summary}
\label{sec:6}

An experimental method has been developed to elucidate the longitudinal dynamics of the harmonic flow, in particular the possible effects of rapidity fluctuation and event-plane decorrelation. This method selects events based on the angle difference, $\Psi_m^{\mathrm {cut}}$, between the $m^{\mathrm{th}}$-order event planes in the forward and backward rapidity, and then measures the rotation of the $n^{\mathrm{th}}$-order EP angle $\Phi_n$ as a function of $\eta$ near the mid-rapidity. This ``event-shape twist'' procedure allows us to distinguish between two competing mechanisms for the rapidity decorrelation: a systematic rotation versus a random fluctuation of event-plane angles along the $\eta$ direction. The former mechanism is expected to lead to, on an event-by-event bases, a non-zero $\eta$- or $\Delta\eta$-dependent sine components in the single particle azimuthal distribution or in the two-particle angular correlations. These non-zero sine components can be used to determine the rotation angle, whose sign and magnitude are fixed by the twist procedure.

The robustness of the event-shape twist technique is demonstrated and the magnitude of the two decorrelation effects is predicted using the AMPT transport model, which is known to contain significant longitudinal fluctuations and EP decorrelation effects~\cite{Xiao:2012uw,Pang:2012he,Pang:2012uw}. A significant rotation of $\Phi_n$ is observed near mid-rapidity for events selected to have a large twist angle $\Psi_m^{\mathrm {cut}}$ for $m=2$ and 3, and the rotation in $\Phi_n$ is observed to vary linearly with $\eta$. This rotation is observed not only for $n=m$ but also for $n>m$. For example, a significant rotation is observed in $\Phi_4$ and $\Phi_5$ for events selected on the $\Psi_2^{\mathrm {cut}}$, as well as in $\Phi_5$ for events selected on the $\Psi_3^{\mathrm {cut}}$. This behavior is consistent with the effects of non-linear coupling between $v_n$ of different order, i.e. the coupling of $v_4$ and $v_5$ to $v_2$ and $v_5$ to $v_3$. Furthermore, a significant fraction of the observed rapidity decorrelation in the AMPT model is found to arise from a systematic rotation of EP angles along the $\eta$ direction, the remaining fraction is consistent with a random fluctuation of EP angle in $\eta$. 

The results obtained in this study are qualitatively consistent with hydrodynamic response to the initial state fireball that is twisted along rapidity, as proposed in Ref.~\cite{Bozek:2010vz}; Our later study published in a separate paper~\cite{Jia:2014ysa} traces this twist to the independent fluctuations of the eccentricity vector for the projectile nucleus and the eccentricity vector for the target nucleus. This twist intrinsically a long-range effect despite of the apparent breaking of the factorization relation Eq.~\ref{eq:flow2}, and it naturally predicts a decrease of the ridge amplitude and broadening of the ridge width at large $\Delta\eta$ as shown in Fig.~\ref{fig:8} (b), which can be measured experimentally. Furthermore, if the twist effect dominates the longitudinal flow angle fluctuation, then one would expect larger signal in p+A collisions and peripheral A+A collisions where the FB-asymmetry is bigger, as well as in lower collision energy at RHIC where the system is less boost-invariant.

The results in this paper focus on the top 10\% of the events with largest twist. Results in other event classes show smaller twist and relatively larger random fluctuation contribution. However, since the average twist can always be extracted for each event class, we can always statically separate the two contributions. Note that the random fluctuation component of EP decorrelation could be related to other initial and final state effects, such as initial flow~\cite{Pang:2012he} and hydrodynamic noise~\cite{Murase:2013tma}, these effects can and should be investigated more quantitatively in model calculations.

The event-shape twist technique is a promising tool for studying the longitudinal dynamics of flow fluctuations, in particular for understanding the origin of the event-plane decorrelation and for quantifying the factorization of the harmonic coefficients of the two-particle angular correlations into a product of single-particle flow coefficients. The main advantage of the technique is to select preferably events with large twist angle, so the signal remain large after the ensemble average. This method can be easily implemented in the experimental data analysis, as well as theoretical calculations. 

We appreciate valuable comments from R.~Lacey. This research is supported by NSF under grant number PHY-1305037 and by DOE through BNL under grant number DE-AC02-98CH10886.
\bibliography{twist}{}
\bibliographystyle{apsrev4-1}
\end{document}